\documentclass[twocolumn,superscriptaddress,prl]{revtex4-1}
\usepackage{mathrsfs,braket}
\usepackage{amssymb, amsbsy, amsmath, latexsym, dsfont, array, layout,
graphicx,mathrsfs,color,ulem,bm}

\begin{document}

\title{Topological quantum walks in momentum space with a Bose-Einstein condensate}%

\author{Dizhou Xie}
\thanks{These authors contributed equally to this work}
\affiliation{
Interdisciplinary Center of Quantum Information, State Key Laboratory of Modern Optical Instrumentation, and Zhejiang Province Key Laboratory of Quantum Technology and Device of Physics Department, Zhejiang University, Hangzhou 310027, China}
\author{Tian-Shu Deng}
\thanks{These authors contributed equally to this work}
\affiliation{CAS Key Laboratory of Quantum Information, University of Science and Technology of China, Hefei 230026, China}
\author{Teng Xiao}
\affiliation{
Interdisciplinary Center of Quantum Information, State Key Laboratory of Modern Optical Instrumentation, and Zhejiang Province Key Laboratory of Quantum Technology and Device of Physics Department, Zhejiang University, Hangzhou 310027, China}
\author{Wei Gou}
\affiliation{
Interdisciplinary Center of Quantum Information, State Key Laboratory of Modern Optical Instrumentation, and Zhejiang Province Key Laboratory of Quantum Technology and Device of Physics Department, Zhejiang University, Hangzhou 310027, China}
\author{Tao Chen}
\affiliation{
Interdisciplinary Center of Quantum Information, State Key Laboratory of Modern Optical Instrumentation, and Zhejiang Province Key Laboratory of Quantum Technology and Device of Physics Department, Zhejiang University, Hangzhou 310027, China}
\author{Wei Yi}
\email{wyiz@ustc.edu.cn}
\affiliation{CAS Key Laboratory of Quantum Information, University of Science and Technology of China, Hefei 230026, China}
\affiliation{CAS Center For Excellence in Quantum Information and Quantum Physics, Hefei 230026, China}
\author{Bo Yan}%
\email{yanbohang@zju.edu.cn}
\affiliation{
Interdisciplinary Center of Quantum Information, State Key Laboratory of Modern Optical Instrumentation, and Zhejiang Province Key Laboratory of Quantum Technology and Device of Physics Department, Zhejiang University, Hangzhou 310027, China}
\affiliation{
Collaborative Innovation Centre of Advanced Microstructures, Nanjing University, Nanjing, China, 210093}
\affiliation{
Key Laboratory of Quantum Optics, Chinese Academy of Sciences, Shanghai 200800, China}

\date{\today}

\begin{abstract}
We report the experimental implementation of discrete-time topological quantum walks of a Bose-Einstein condensate in momentum space.
Introducing stroboscopic driving sequences to the generation of a momentum lattice, we show that the dynamics of atoms along the lattice is effectively governed by a periodically driven Su-Schrieffer-Heeger model, which is equivalent to a discrete-time topological quantum walk.
We directly measure the underlying topological invariants through time-averaged mean chiral displacements, which are consistent with our experimental observation of topological phase transitions. We then observe interaction-induced localization in the quantum-walk dynamics, where atoms tend to populate a single momentum-lattice site under interactions that are non-local in momentum space. Our experiment opens up the avenue of investigating discrete-time topological quantum walks using cold atoms, where the many-body environment and tunable interactions offer exciting new possibilities.
\end{abstract}
\maketitle

Exploring topological phases is a main theme in modern physics. Characterized by topological invariants which reflect the global geometric properties of the system wave function, topological phases host a range of fascinating features, which are robust to local perturbations and potentially useful for applications in quantum information~\cite{HKrmp10,QZrmp11}. Besides conventional topological materials in solid-state systems, topological phenomena also emerge away from equilibrium. For example, topological phenomena exist in non-Hermitian open systems~\cite{RL09,ESHK11,LH13,chenshu,Lee,Wang1,Wang2,Kunst18,Uedaprx,Ueda18,LeeJY,Dasreview}, in quench processes and periodically driven Floquet systems~\cite{Demler10,Zoller11,Levin13,Gogolin15,Bhaseen15,rigol,goldman,Vishwanath16,Sondhi16,Heyl13,BH16,Balatsky,Budich16,Refael16,Zhai17,Chen17,Ueda17,xiongjun,wyi}, which have stimulated intense interest recently due to the rapid progress in synthetic quantum-simulation platforms such as cold atoms~\cite{ETHcoldatom14,Weitenberg2016,Weitenberg17,chenshuai}, photonics~\cite{KB+12,Cardano2016,Cardano2017,BNE+17,Weimannnm,PTsymm2,Zeunerprl,pxprl,pxdqpt,pxchern,chuanfengprl,chuanfengdqpt,chaoyangprl,xiangdongprl}, phononics~\cite{Chen2018}, and superconducting qubits~\cite{dongning}.

In particular, photonic topological quantum walks have proved to be a versatile platform for investigating topological phenomena in both unitary~\cite{KB+12,Cardano2016,Cardano2017,BNE+17,chuanfengprl,chaoyangprl,chuanfengdqpt,xiangdongprl} and non-unitary~\cite{PTsymm2,Zeunerprl,pxprl,pxdqpt,pxchern} Floquet dynamics. In cold atoms, whereas Floquet topological phases~\cite{ETHcoldatom14} and quantum walks~\cite{becqw} have been respectively implemented, quantum walks with topological properties
are yet to be realized. In contrast to photonic quantum walks, where only dynamics of up to two strongly correlated photons have been reported~\cite{entangleqw,qubitqw}, the quantum-many-body nature and the tunable interactions~\cite{FRreview} of cold atoms offer the exciting possibility of exploring topological quantum walks in the presence of many-body entanglement~\cite{twinfock} or strong interactions.

Here we report the experimental implementation of discrete-time topological quantum walks in momentum space for a Bose-Einstein condensate (BEC). Building upon the technique of momentum-lattice generation~\cite{ml1,ml2,ml3,ml4,ml5}, we introduce a staggered, time-periodic driving sequence to the Raman-induced tunneling along the lattice, such that dynamics of the condensate atoms is well-described by
a discrete-time quantum walk which supports Floquet topological phases. Our scheme is in contrast to previous implementations of photonic quantum walks, where the dynamics is driven by the propagation of classical light or photons rather than by Hamiltonians.

We confirm topological properties of the atomic quantum walk by measuring dynamic signatures such as time-averaged mean chiral displacement (A-MCD)~\cite{Cardano2017} and second-statistical moment~\cite{Cardano2016}. As an illustrative example of the tunability of cold atoms, we experimentally demonstrate the interaction-induced localization in the quantum-walk dynamics, where the degree of localization sensitively depends on the BEC density.

\begin{figure}[tbp]
\includegraphics[width= 0.47\textwidth]{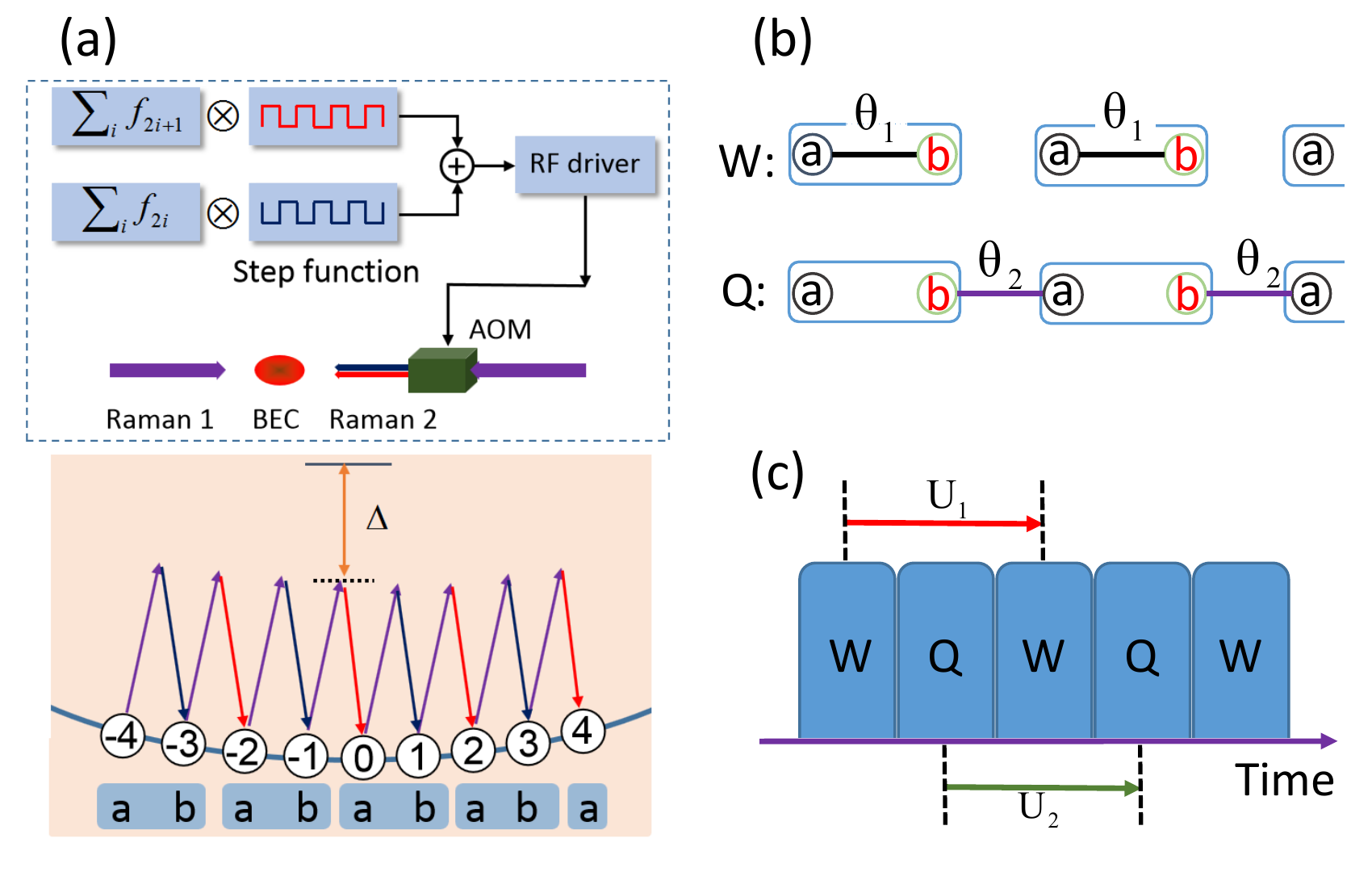}
\caption{\label{scheme}(Color online) Experimental implementation of momentum-space discrete-time quantum walks with cold atoms. (a) Atoms in the BEC are coupled by counter-propagating Raman beams. One of the Raman beams (Raman 2) is imprinted with multiple-frequency components via an acoustic-optical modulator (AOM). The odd and even frequency components, marked by red and blue arrows, alternate in time, which are controlled by the radio-frequency (RF) driver. The periodically modulated Raman processes couple
discretized momentum states, leading to a momentum lattice labeled by $n\in\mathbb{Z}$ with nearest-neighbor Raman-assisted hopping. The staggered Raman couplings enable us to define sublattice sites $a$ and $b$ along the lattice.
(b) Illustration of the effective time-evolution operators $W$ and $Q$. Under $W$ ($Q$), the intra-cell (inter-cell) couplings are turned on along the momentum lattice. (c) Illustration of the two time frames, dictated by Floquet operators $U_1$ and $U_2$ (see main text), respectively.}
\label{fig:setup}
\end{figure}

{\it Discrete-time quantum walks in momentum space:-}
We implement discrete-time quantum walks in momentum space using Raman-induced momentum lattice with periodic driving (see Fig.~\ref{fig:setup}). In previous experiments, momentum lattices have been realized for cold atoms, thanks to the precise control of momentum states with multi-frequency Raman beams~\cite{ml1,ml2,ml3,ml4,ml5}. Here we further introduce periodic switching of the odd- and even-frequency components of the multi-frequency Raman lasers [see Fig.~\ref{fig:setup}(a)].
This gives rise to a stroboscopic switching of the hopping terms between adjacent sites along the momentum lattice. Here the $n$-th site ($n\in \mathbb{Z}$) along the lattice corresponds to the momentum $p_n=n\times2\hbar k$, where $k$ is the wave vector of the Raman lasers. As illustrated in Fig.~\ref{fig:setup}(a), atoms on site $n$ and $n+1$
are coupled by a pair of Raman beams with an offset frequency $f_n=(2n+1)\times4E_r/h$, where $E_r$ is the recoil energy.
In order to realize the switching of Raman couplings, the offset frequencies are divided into odd and even components depending on the parity of $n$. While these odd- and even-frequency components are switched on and off by step functions through the RF driver, the effective Hamiltonian for the BEC is given by a periodically-driven Su-Schrieffer-Heeger model
\begin{align}
{\hat H}=\sum\limits_{m}\left[w(t)\left| m,b \right\rangle \left\langle m,a \right| + q(t)\left| m+1,a \right\rangle \left\langle m,b \right| + H.c.\right],\label{eq:ssh}
\end{align}
where non-resonant Raman couplings, corresponding to long-range hoppings along the lattice, are neglected. {This would be a good approximation for $\hbar\Omega\ll 4E_r$, whereas typical experimental parameters feature $\hbar\Omega\sim E_r$. While these non-resonant couplings are a major source of imperfection, they do not qualitatively change our results~\cite{supp}.}
Here $m$ labels the unit cell, and $a$ and $b$ are the sublattice sites, with $|m,a\rangle$ ($|m,b\rangle$) corresponding to the momentum-lattice site $n=2m$ ($n=2m+1$).
Importantly, the step-wise Raman-induced hopping rates $w(t)$ and $q(t)$ satisfy $w(t)+q(t)=-\hbar\Omega/2$, where $\Omega$ is the Raman-coupling rate and $w(t)$ is given by
\begin{align}
w(t)=\begin{cases}
-\frac{\hbar\Omega}{2},& jT<t\leq jT+t_w\\
0,& jT+t_w<t\leq (j+1)T
\end{cases}.
\end{align}
Here $j\in\{0,1,2,....\}$, $T=t_w+t_q$, and $t_w$ and $t_q$ are pulse durations for Raman processes with even- and odd-frequency components, respectively.

\begin{figure}[tbp]
\hspace{-0.5cm}
\includegraphics[width=9cm]{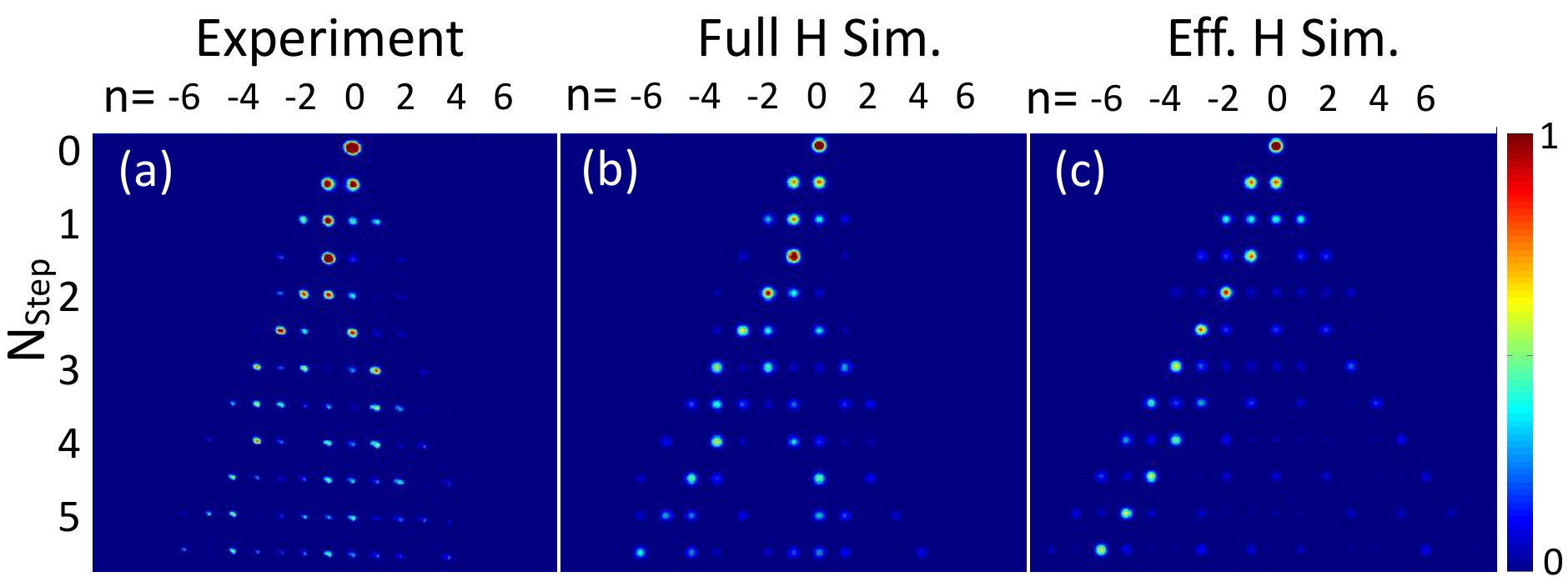}
\caption{(Color online) (a) Experimental demonstration of a typical discrete-time quantum-walk dynamics in momentum space. The color scale indicates atomic population of the corresponding momentum-lattice sites, normalized by the total atomic population detected at each time step. (b)(c) Numerical simulation of the atomic population at each time step driven by (b) the full Hamiltonian~\cite{supp}, and (c) the effective Hamiltonian (1), respectively. For both the experiment and the numerical simulations, the atoms are initialized in the ground state of the BEC with $n=0$, and are evolved under $(\theta_1,\theta_2)=(\pi/4,\pi/4)$. Deviations of the experimental measurements from numerical simulations are due to experimental decoherence~\cite{footnote}.}
\label{fig:fig2new}
\end{figure}

Dynamics under Eq.~(\ref{eq:ssh}) can be mapped to discrete-time quantum-walk dynamics governed by Floquet operators. As illustrated in Fig.~\ref{fig:setup}(b), when the even-frequency Raman coupling is turned on, intra-cell hopping $w(t)$ in Eq.~(\ref{eq:ssh}) is finite while $q(t)=0$. It follows that the corresponding time-evolution operator $W(\theta_1)$ is
\begin{align}
W(\theta_1)\left| m,a(b) \right\rangle=\cos\theta_1\left| m,a(b) \right\rangle+i \sin\theta_1\left| m,b(a) \right\rangle,\nonumber\\
\end{align}
where $\theta_1=\Omega t_w/2$. Similarly, when the odd-frequency Raman coupling is turned on, inter-cell hopping $q(t)$ is finite and $w(t)=0$. The time-evolution operator $Q(\theta_2)$ is
\begin{align}
Q(\theta_2)\left| m,a (b) \right\rangle=\cos\theta_2\left| m,a(b) \right\rangle+i\sin\theta_2\left| m\mp 1,b(a) \right\rangle,\nonumber\\
\end{align}
where $\theta_2=\Omega t_q/2$. Note that the condition $t_w+t_q=T$ translates to $\theta_1+\theta_2=\Omega T/2$.
The overall dynamics is thus governed the Floquet operator $U=Q(\theta_2)W(\theta_1)$, which drives the discrete-time quantum walk.
We note that the static effective Hamiltonian $H_{\rm F}$ associated with the Floquet operator is defined through $U=e^{-iH_{\rm F}T/\hbar}$, which is different from Hamiltonian (1).

Experimentally, we prepare a $^{87}$Rb BEC with $\sim 1\times10^5$ atoms in a crossed dipole trap with trapping frequencies $2\pi\times(115,40,100)$Hz. We generate the momentum lattice with a pair of multi-frequency Raman lasers at $1064$nm, following the procedure outlined in Ref.~\cite{ml5}.
Quantum walks are introduced by a periodical modulation of the Raman pulses.
For detection, we turn off the dipole trap and Raman beams, and take an absorption image after $20$ms time of flight, from which atomic populations at different momenta are extracted. For all our experiments, we set $\Omega=2\pi\times2.3(1)$kHz and $E_r/\hbar =2\pi\times2.03$kHz, while $\theta_1$ and $\theta_2$ are tuned by adjusting pulse durations $t_w$ and $t_q$.

A typical experimental measurement for a homogeneous discrete-time quantum-walk dynamics is shown in Fig.~\ref{fig:fig2new}, where a ballistic spreading of population, typical for discrete-time quantum walks, is observed. Here we fix $T\approx 0.22$ms and $\theta_1=\theta_2=\pi/4$.
The experimental observation agrees well with numerical simulations using either the full Hamiltonian (with non-resonant Raman terms) or the effective Hamiltonian (1), especially at short times. At long times, deviations become manifest due to decoherence from a range of sources~\cite{footnote}.

\begin{figure}[tbp]
\includegraphics[width= 0.47\textwidth]{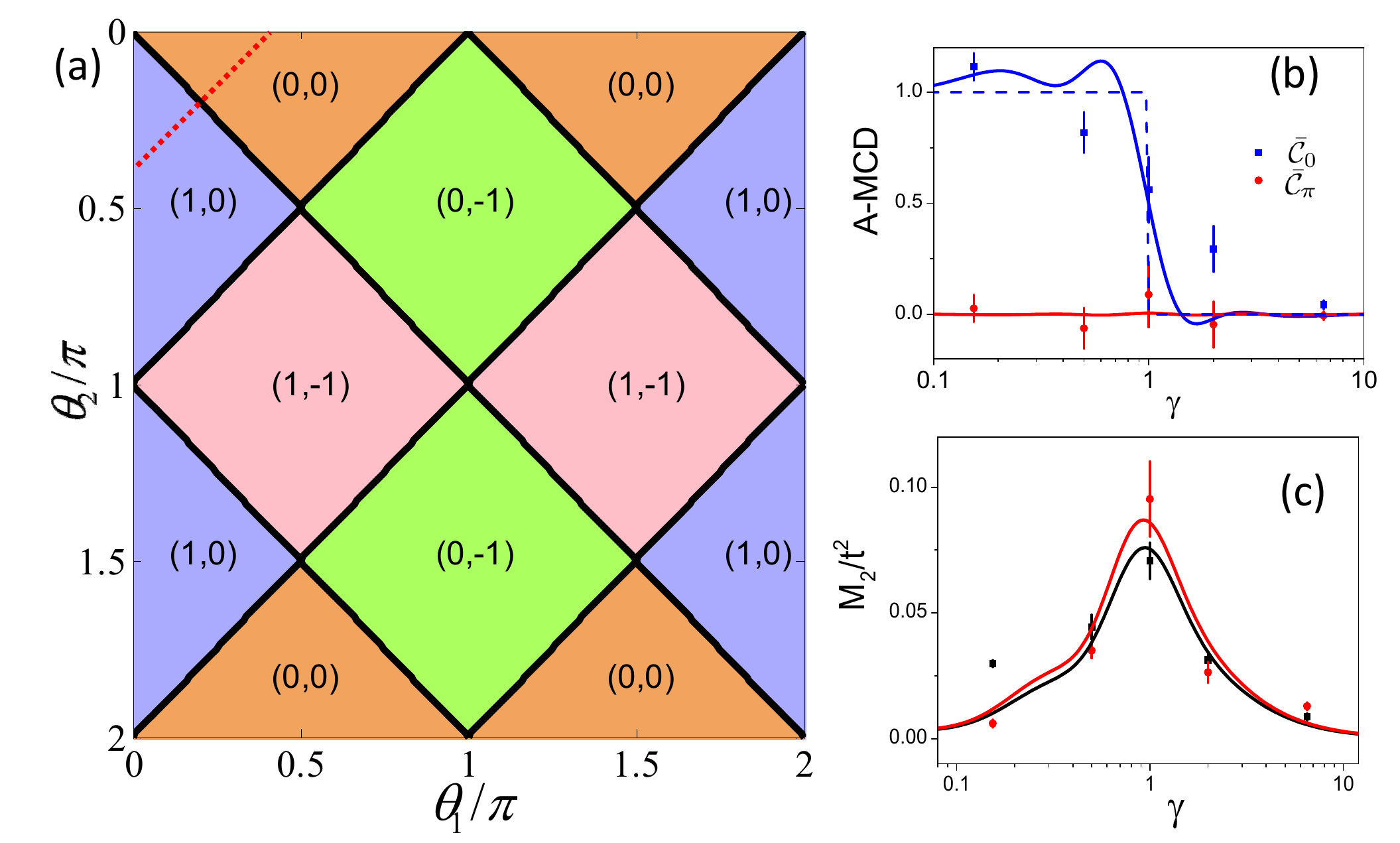}
\caption{(Color online) (a) Topological phase diagram of our discrete-time quantum walk. Topological invariants $(C_0,C_\pi)$ are shown on the plane of $(\theta_1,\theta_2)$. The dashed red line indicates the parameters traversed in (b)(c).
(b) Experimentally measured A-MCD for a six-step quantum walk, where each data point is averaged over $3$ measurements.
The dashed blue (red) line shows the variation of $C_0$ ($C_\pi$). (c) Experimentally measured $M^{(\alpha)}_2/N_{\rm step}^2$ for dynamics governed by $U_1$ (black) and $U_2$ (red), respectively. Each data point is averaged over $11$ measurements. In (b)(c), the experimental data (dots with error bars) agree well with results from numerical simulations (solid lines) using the effective Hamiltonian (1). Error bars in (b)(c) reflect the standard deviations.}\label{fig:fig3}
\end{figure}

{\it Detecting topological properties:-}
Quantum walks governed by $U=Q(\theta_2)W(\theta_1)$ support Floquet topological phases, which are characterized by a pair of winding numbers
defined in distinct time frames~\cite{AO13}. As illustrated in Fig.~\ref{fig:setup}(c), these time frames are associated with the Floquet operators
\begin{align}
U_1&=W(\frac{\theta_1}{2})Q(\theta_2)W(\frac{\theta_1}{2}),\nonumber\\
U_2&=Q(\frac{\theta_2}{2})W(\theta_1)Q(\frac{\theta_2}{2}),
\end{align}
which have chiral symmetry $\Gamma U_{\alpha}\Gamma^{-1}=U^{-1}_{\alpha}$ ($\alpha=1,2$) with $\Gamma=|a\rangle\langle a|-|b\rangle\langle b|$,
and give rise to winding numbers $C_1$ and $C_2$, respectively. We note that the non-resonant Raman couplings neglected in (1) can break chiral symmetry and give rise to experimental error~\cite{supp}. However, these terms are far-detuned with two-photon detunings much larger than $\Omega$, such that topological features of the quantum-walk dynamics are still manifest in our experimental data.

Following Ref.~\cite{AO13}, we define topological invariants $(C_0,C_\pi)=(\frac{C_1+C_2}{2},\frac{C_1-C_2}{2})$, which dictate the number of topological edge states with quasienergies $ET/\hbar=0$ and $ET/\hbar=\pi$, respectively, through the bulk-boundary correspondence. The topological phase diagram of the system is shown in Fig.~\ref{fig:fig3}(a), where $(C_0,C_\pi)$ are plotted as functions of the Raman-coupling parameters $(\theta_1,\theta_2)$.

To experimentally demonstrate topological features of the momentum-space quantum walk, we experimentally probe $(C_0,C_\pi)$, and confirm topological phase transitions by monitoring dynamics of condensate atoms in momentum space.
Quantum walks in different time frames are implemented by applying different sequence of Raman pulses which correspond to $U_1$ and $U_2$, respectively. Here we fix $T\approx 0.16$ms, and adjust $t_w$ and $t_q$ so that $\theta_1+\theta_2=3\pi/8$, as shown by the red dashed line in Fig.~\ref{fig:fig3}(a).

First, we directly probe topological invariants $(C_0,C_\pi)$ by detecting chiral displacements~\cite{Cardano2017}. We initialize atoms in the $n=0$ state, let them evolve on the lattice, and take time-of-flight images at different time steps.
Interestingly, we fin that A-MCD converges much faster to the topological invariant than the commonly sued mean chiral displacement~\cite{supp}.
For an $N$-step quantum walk, the A-MCD is defined as
\begin{align}
\bar{\mathcal{C}}_{\alpha}=\frac{2}{N}\sum_{N_{\rm step}=1}^{N}\sum_m m\left[P^{(\alpha)}_{m,a}(N_{\rm step})-P^{(\alpha)}_{m,b}(N_{\rm step})\right],
\end{align}
where $\alpha\in\{1,2\}$ indicates the time frame, and $P^{(\alpha)}_{m,a(b)}(N_{\rm step})$ is the measured atom population in the state $|m,a(b)\rangle$ at the $N_{\rm step}$-th step ($N_{\rm step}\in \mathbb{N}$) for the dynamics under $U_\alpha$.
Performing the measurements in both time frames, we construct $ \bar{\mathcal{C}}_{0,\pi}=\frac{1}{2}(\bar{\mathcal{C}}_{1}\pm \bar{\mathcal{C}}_{2})$, which should approach $C_{0,\pi}$ in the long-time limit.
In our experiment, as shown in Fig.~\ref{fig:fig3}(b), the measured A-MCDs agree well with theoretical predictions after a six-step quantum walk.

An important observation of the measured A-MCD is the occurrence of a topological phase transition near $\gamma=1$ ($\gamma=\theta_1/\theta_2$), where the numerically calculated winding numbers $C_1$ and $C_2$ undergo abrupt changes. To confirm this, we measure the second-statistical moment $M^{(\alpha)}_2$ characterizing the probability distribution of the walker position, which is defined as
\begin{align}
M^{(\alpha)}_2=\sum_m m^2 \left[P^{(\alpha)}_{m,a}(N_{\rm step})+P^{(\alpha)}_{m,b}(N_{\rm step})\right]
\end{align}
for the $N_{\rm step}$-th step.
In the long-time limit, $M^{(\alpha)}_2/N_{\rm step}^2$ should peak at the topological phase boundary under $U_{\alpha}$~\cite{Cardano2016}, where the corresponding winding number undergoes an abrupt jump.
In Fig.~\ref{fig:fig3}(c), we show measured $M^{(\alpha)}_2/N_{\rm step}^2$ after six time steps. The measured peaks in $M^{(\alpha)}_2/N_{\rm step}^2$ for both time frames are located near $\gamma=1$, consistent with theoretical predictions. The measured topological invariants and phase transitions are also consistent with edge-state measurement when an open boundary is imposed~\cite{supp}.

{\it Interaction-induced localization:-}
A key advantage of an atomic quantum-walk platform is the highly tunable many-body environment and interactions. As an illustrative example, we study the impact of interactions on our quantum-walk dynamics. We consider the short-range, $s$-wave interactions between $^{87}$Rb atoms, which translate to non-local interactions on the momentum lattice~\cite{ml2}. Under the Hartree-Fock approximation and taking the mean-fields of BEC atoms on each momentum-lattice site, the equations of motion of the system during an interacting quantum-walk dynamics can be written as~\cite{supp}
\begin{align}
\label{CPeq}
i\hbar\frac{\text{d}}{\text{d}t}\Phi=\bar{H}\Phi,
\end{align}
where $\Phi=[\phi_{m,a},\phi_{m,b}]^{T}$ with $\phi_{m,a}$ ($\phi_{m,b}$) representing the mean-field wave function on the momentum-lattice site $|m,a\rangle$ ($|m,b\rangle$), with the normalization condition $\sum_{m,\sigma}|\phi_{m,\sigma}|^2=1$. The matrix elements of $\bar{H}$ are ($\sigma=a,b$)
\begin{align}
&\bar{H}_{m,a;m,b}=\bar{H}_{m,b;m,a}=w(t)\\
&\bar{H}_{m+1,a;m,b}=\bar{H}_{m,b;m+1,a}=q(t)\\
&\bar{H}_{m,\sigma;m,\sigma}=U|\phi_{m,\sigma}|^{2}+2U\sum_{m'\neq m,\sigma'}|\phi_{m',\sigma'}|^{2}.\label{eq:int}
\end{align}
Here $a_s$ is the $s$-wave scattering length, $\mu$ is the atomic mass, the interaction energy $U=g\rho$, where $g=4\pi\hbar^2a_s /\mu$, and $\rho$ is the overall BEC density. In deriving Eq.~(\ref{CPeq}), we use the effective Hamiltonian (\ref{eq:ssh}) in addition to the interaction terms~\cite{supp}.

\begin{figure}[tbp]
\includegraphics[width= 8.5cm]{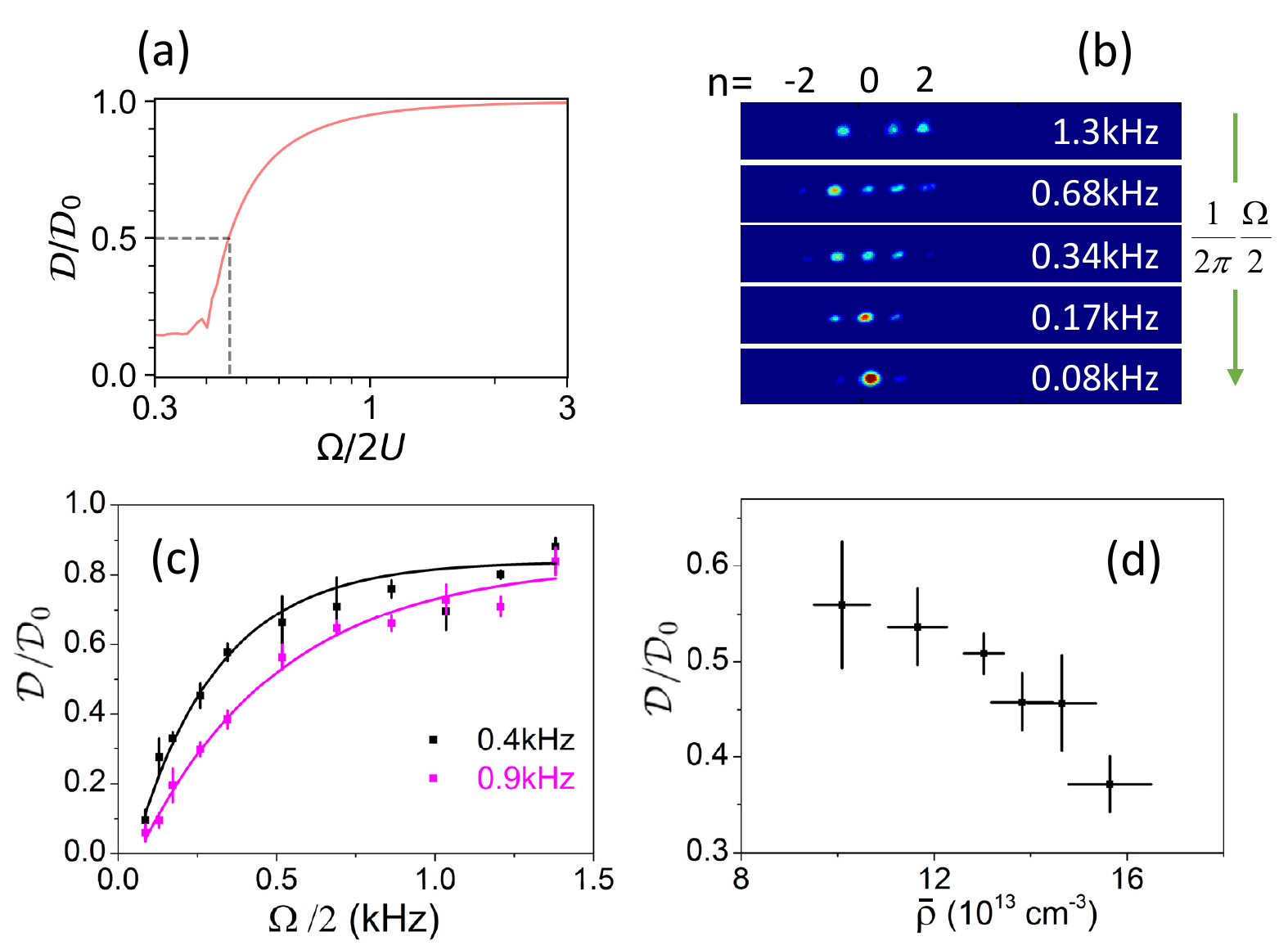}
\caption{(Color online) Interaction-induced localization. (a) Numerically simulated normalized mean distance $\mathcal{D/D}_0$ as a function of $\hbar\Omega/2U$ after an eight-step quantum walk.  (b) Typical time-of-flight images for two-step quantum walks, with $U/\hbar=2\pi\times0.9(1)$kHz. From top to bottom, the Raman-induced tunneling rate $\Omega/2$ varies from $2\pi\times1.3$kHz to less than $2\pi\times100$Hz. (c) $\mathcal{D}/\mathcal{D}_0$ versus $\Omega/2$ for $U/\hbar=2\pi\times0.41(6)$kHz (black) and $2\pi\times0.9(1)$kHz (red). (d) Density dependent $\mathcal{D}/\mathcal{D}_0$. The coupling strength is $\Omega/2=2\pi\times0.35$kHz. The BEC density is decreased by holding BEC in the dipole trap for a long time. The trap frequencies of dipole trap are $2\pi\times(214, 61, 220)$Hz.
}\label{fig:fig4}
\end{figure}

A direct consequence of interaction, as manifested in Eq.~(\ref{eq:int}), is that the interaction energy between atoms on the same momentum-lattice site is half of that between atoms from different sites. This is due to the presence of an exchange term for two interacting atoms in distinct momentum states. For an interacting BEC, such an effect gives rise to localization of atoms during the quantum-walk dynamics, since the difference in the interaction-energy shift plays the role of an effective detuning which hinders Raman-induced tunnelings.
To see the effect, we numerically evolve Eq.~(\ref{CPeq}), assuming a constant $U$~\cite{supp}, up to the time of an eight-step quantum walk.
In Fig.~\ref{fig:fig4}(a), we plot the numerically calculated mean distance $\mathcal{D}$, defined through
$\mathcal{D}=\sum_m (|2m||\phi_{m,a}|^2+|2m+1||\phi_{m,b}|^2)$, as a function of $\hbar\Omega/2U$.
With decreasing $\hbar\Omega/2U$, interactions dominate the process and $\mathcal{D}$ approaches zero, indicating the localization of atoms in their initial state. In contrast, with increasing $\hbar\Omega/2U$, couplings dominate the process and $\mathcal{D}$ asymptotically approaches $\mathcal{D}_0$, the mean distance for the non-interacting case with $U=0$. Importantly, a localization-delocalization transition can be identified near $\hbar\Omega/2U\simeq0.45$.

To experimentally study the interaction effects, we characterize the interaction energy by $U=g\bar\rho$, where $\bar\rho$ is the average density of the initial BEC prior to quantum-walk dynamics evaluated with the Thomas-Fermi radius. We first vary $\hbar\Omega$ at a fixed $U/\hbar=2\pi\times0.9(1)$kHz. We evolve the BEC under the Floquet operator $U_1$ with $\theta_1=\theta_2=3\pi/16$ for two steps, before we take a time-of-flight image of the atoms.
Figure~\ref{fig:fig4}(b) shows a typical image for various values of $\Omega/2$ ranging from $2\pi\times1.3$kHz to less than $2\pi\times100$Hz. Apparently, the atoms become more and more localized at their initial lattice site $|0,a\rangle$ as the rate $\hbar\Omega/2U$ decreases.
In Fig.~\ref{fig:fig4}(c), we show the measured the normalized mean distance $\mathcal{D}/\mathcal{D}_0$ as a function of $\Omega/2$ for different $U$. Whereas both curves are more smooth compared to the numerically-simulated curve in Fig.~\ref{fig:fig4}(a), our experimental results are consistent with the numerical prediction that a localization-delocalization transition occurs near $\hbar\Omega/2U\simeq0.45$. The difference between Figs.~\ref{fig:fig4}(a) and (c) originates from the finite time of the quantum-walk dynamics, as well as from the spatially non-uniform $U$ in our experiment.

To confirm that the observed localization in the quantum-walk dynamics is unequivocally induced by interactions, we vary $\bar\rho$ by holding the BEC in the dipole trap for different durations, prior to the quantum-walk experiment. This allows us to directly examine the density dependence of the localization with a fixed $\Omega/2=2\pi\times0.35(2)$kHz. As shown in Fig.~\ref{fig:fig4}(d), the normalized mean distance $\mathcal{D}/\mathcal{D}_0$ sensitively depends on the BEC density at the trap center, with increasing localization for larger densities.
We therefore conclude that we have indeed observed the interaction-induced localization in quantum-walk dynamics of a BEC in the momentum space.
Finally, we note that for our experiments characterizing topological features, $U/\hbar\approx 2\pi\times0.6(1)$kHz and $\hbar\Omega/2U\simeq2$, such that the localization effect due to interactions is negligible. In the future, it would be interesting to explore the interplay of topology and interaction bases on our platform.

{\it Discussion:-}
We experimentally implement a stroboscopic driving of ultracold atoms on a momentum lattice, thus realizing discrete-time quantum-walk dynamics using cold atoms. The accessible number of time steps in our experiment is affected by the finite size and inhomogeneous density distribution of the BEC in the trapping potential. Those inhomogeneities leads to broadening and inhomogeneous interactions in momentum space, which give rise to decoherence in the dynamics. A practical solution is to further weaken the trapping potential by loading the BEC into a gradient magnetic field which offsets gravity.
Our scheme offers exciting possibilities of exploring topological quantum-walk dynamics in the context of strongly-correlated many-body systems with tunable interatomic interactions or strong many-body entanglement~\cite{twinfock}.

{\it Acknowledgement:-}
We are grateful to Ying Hu for helpful comments. We acknowledge the support from the National Key R$\&$D Program of China under Grant No.2018YFA0307200, National Natural Science Foundation of China under Grant Nos. 91636104, 11974331 and 91736209, Natural Science Foundation of Zhejiang province under Grant No. LZ18A040001, and the Fundamental Research Funds for the Central Universities. W. Y. acknowledges support from the National Key Research and Development Program of China (Grant Nos. 2016YFA0301700 and 2017YFA0304100).

\newpage
\begin{widetext}
\section {Supplemental Materials for ``Topological quantum walks in momentum space with a Bose-Einstein condensate''}
In this Supplemental Materials, we provide details for the derivation of the effective Hamiltonian and the topological invariants. We also provide additional simulation and experimental data demonstrating time-averaged mean chiral displacement, the detection of topological edge states, the comparison between the full and effective Hamiltonians, and the derivation of Eq.~(8) in the main text.

\subsection{Derivation of the effective Hamiltonian}
In this section, we derive, from the full Hamiltonian, the effective Hamiltonian as shown in Eq.~(1) of the main text.
Following the experimental setup illustrated in Fig.~1 of the main text, we start from the single-particle Hamiltonian under the dipole approximation
\begin{equation}
 {\hat H}=\frac{{\hat{\bf p}}^2}{2M}+\hbar\omega_e\left| e \right\rangle \left\langle e \right|+\hbar {\omega _g}\left| g \right\rangle \left\langle g \right|+{\bf d}\cdot{\bf E},
\end{equation}
where $|g\rangle$ ($|e\rangle$) is the atomic ground (excited) state with energy $\hbar\omega_g$ ($\hbar\omega_e$), $M$ is the atomic mass, and ${\bf d}$ is the electric dipole moment.
The electric field is given by ${\bf E}={{\bf E}_+}+{{\bf E}_-}$, with
\begin{align}
{{\bf E}_+}&={{\bf E}_+}\cos({{\bf k}_+}\cdot{\bf x}-{\omega_+}t+{\phi_+}),\\
{{\bf E}_-}&=\sum\limits_i { {{\bf E}_i}\cos({{\bf k}_i}\cdot{\bf x}-{\omega_i}t+{\phi_i})},
\end{align}
where ${\Omega_+}=\left\langle e \right| {\bf d}\cdot{\bf {E_+}}\left| g \right\rangle/\hbar$, ${\Omega_i}=\left\langle e \right| {\bf d}\cdot{\bf {E_i}}\left| g \right\rangle/\hbar$, and $\omega_+$ and $\omega_i$ ($\phi_+$ and $\phi_i$) are the frequencies (phases) of acoustic-optical modulated Raman lasers.
According to our experimental configuration (see Fig.~1 of the main text), we write ${{\bf k}_+}=k{\bf e}_x$, ${{\bf k}_i} \simeq -k{\bf e}_x$, and $\Delta  \equiv {\omega _{eg}} - {\omega _ + } \simeq {\omega _{eg}} - {\omega _i}$ with ${\omega _{eg}} \equiv {\omega _e} - {\omega _g}$. Here ${\bf e}_x$ is the unit vector along the $x$ direction.

Without loss of generality, we assume that the Rabi frequencies $\Omega_+$ and $\Omega_i$ are real, so the Hamiltonian can be written as
\begin{align}
 &\hat{H}=\frac{{\hat{\bf p}}^2}{2M}+\hbar\omega_e\left| e \right\rangle \left\langle e \right|+\hbar {\omega _g}\left| g \right\rangle \left\langle g \right|\nonumber\\
&+\hbar\left[\Omega_+\cos( kx-{\omega_+}t
+{\phi_+})+\sum\limits_i { \Omega_i}\cos(-kx-{\omega_i}t+{\phi_i})\right]\nonumber\\
&\times(\left| e \right\rangle \left\langle g \right|+\left| g \right\rangle \left\langle e \right|).
\end{align}
We then apply the rotating-wave approximation, adiabatically eliminate the excited state $|e\rangle$, and expand the external motion of the ground state in the discretized momentum lattice with $\left| \psi \right\rangle=\sum\limits_n { c_n}{e^{i2nkx}}\left| n \right\rangle \otimes \left| g \right\rangle$ ($n\in \mathbb{Z}$). The resulting effective ground-state Hamiltonian becomes
\begin{equation}
\begin{aligned}
 {\hat H}_{\rm{eff}}&=\sum\limits_n 4n^2E_r \left| n \right\rangle \left\langle n \right| \\&+\frac{\hbar}{4\Delta}\sum\limits_n {\Omega_+ \Omega_n e^{-i(\omega_+-\omega_n)t}e^{i(\phi_+-\phi_n)} \left| n+1 \right\rangle \left\langle n \right| }\\&+\frac{\hbar}{4\Delta}\sum\limits_n {\Omega_+ \Omega_n e^{i(\omega_+-\omega_n)t}e^{-i(\phi_+-\phi_n)} \left| {n} \right\rangle \left\langle n+1 \right| } .
\end{aligned}
\end{equation}
Experimentally, we tune the effective Rabi frequencies of all Raman processes to be of the same magnitude, such that $\Omega=\frac{\Omega_n\Omega_+}{2\Delta}$. We then have
\begin{align}
& {\hat H}_{\rm{eff}}^I=\sum\limits_n
\left({\frac{\hbar\Omega}{2}e^{i(2n+1)\frac{4E_r}{\hbar}t} e^{-i(\omega_+-\omega_i)t}e^{i\varphi_n} \left| n+1 \right\rangle \left\langle n \right| }\right.\nonumber\\
&\left.+{\frac{\hbar\Omega}{2}e^{-i(2n+1)\frac{4E_r}{\hbar}t} e^{i(\omega_+-\omega_i)t}e^{-i\varphi_n} \left| {n} \right\rangle \left\langle n+1 \right| }\right),\label{eq:Heff}
\end{align}
where $\varphi_n=(\phi_+-\phi_n)$, and we have taken the appropriate interaction picture.

The effective Hamiltonian (\ref{eq:Heff}) can be further simplified by neglecting non-resonant terms. Experimentally, we choose $\Omega\ll 4E_r$, such that only resonant terms have significant contribution in the dynamics. The resulting Hamiltonian becomes
\begin{align}
 {\hat H}_{\rm{eff}}^I=\sum\limits_n({\frac{\hbar\Omega}{2}e^{i\varphi_n} \left| n+1 \right\rangle \left\langle n \right| }+H.c.),
\end{align}
which is a tight-binding Hamiltonian on a momentum lattice. In the experiment, we set $\varphi_n=0$ for all $n$.

We now consider the periodic switching of the Raman couplings, due to the step functions imposed by the radio-frequecy (RF) driver. Denoting the pulse durations when the even- (odd-) frequency components are switched on as $t_w$ ($t_q$), we map the effective Hamiltonian to a Su-Schiefer-Heeger (SSH) model in momentum space with periodically driven coefficients
\begin{equation}
{\hat H}=\hbar (w(t)\sum\limits_{m}\left| m,b \right\rangle \left\langle m,a \right| + q(t)\sum\limits_{m}\left| m+1,a \right\rangle \left\langle m,b \right|) + H.c.,\label{eq:s8}
\end{equation}
where the mapping between $|m,a(b)\rangle$ as well as the coefficients $w(t)$ and $q(t)$ are given in the main text.

\subsection{Experimental data for alternative Floquet operator}

In Fig.~2 of the main text, we show typical quantum-walk dynamics in momentum space, governed by the Floquet operator $U'=Q(\theta_2)W(\theta_1)$. Here we show additional experimental data for quantum-walk dynamics governed by the Floquet operator $U=W(\theta_1)Q(\theta_2)$. In Fig.~\ref{fig:SUprime}, we see that under essentially the same parameters $(\theta_1,\theta_2)=(\pi/4,\pi/4)$, atoms tend to spread to positive momenta, in the opposite direction compared to that in Fig.~2 of the main text.

\begin{figure*}[tbp]
\includegraphics[width= 0.8\textwidth]{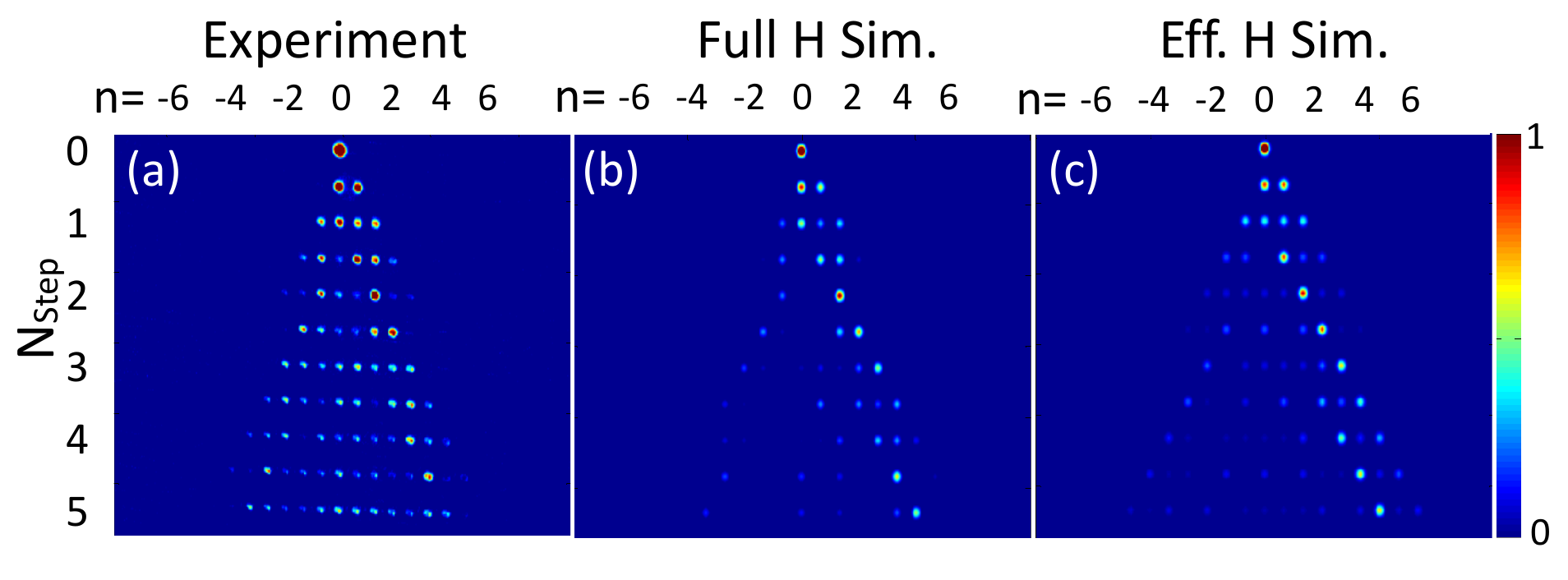}
\caption{(Color online) (a) Experimental demonstration of a typical discrete-time quantum-walk dynamics in momentum space under the Floquet operator $U'$. The color scale indicates atomic population of the corresponding momentum-lattice sites, normalized by the total atomic population detected at each time step. (b)(c) Numerical simulation of the atomic population at each time step driven by (b) the full Hamiltonian Eq.~(\ref{eq:Heff}), and (c) the effective Hamiltonian Eq.~(\ref{eq:s8}), respectively. For both the experiment and the numerical simulations, the atoms are initialized in the ground state of the BEC with $n=0$, and are evolved under $(\theta_1,\theta_2)=(\pi/4,\pi/4)$ with $\Omega=2\pi\times 2.3$kHz.}
\label{fig:SUprime}
\end{figure*}

\subsection{\label{sec:level1}Topological invariants of discrete-time quantum walks}
The discrete-time quantum walk governed by $U=Q(\theta_2)W(\theta_1)$ supports Floquet topological phases, whose topological invariants can be calculated by going to different time frames. These time frames are respectively governed by the Floquet operators
\begin{align}
U_{1} &=W(\frac{\theta_1}{ 2}) Q(\theta_2) W(\frac{\theta_1 }{ 2}), \\
U_{2} &=Q(\frac{\theta_2}{ 2}) W(\theta_1) Q(\frac{\theta_2}{ 2}).
\end{align}

The topology of these Floquet operators are characterized by their associated winding numbers $(C_1,C_2)$. To calculate the winding numbers, we perform Fourier transforms on $U_1$ and $U_2$, and write their Fourier components as
\begin{align}
U_{1} &=d_{0}^{(1)} \sigma_0-i d_{1}^{(1)} \sigma_{x}-i d_{2}^{(1)} \sigma_{y}-i d_{3}^{(1)} \sigma_{z}, \\
U_{2} &=d_{0}^{(2)} \sigma_0-i d_{1}^{(2)} \sigma_{x}-i d_{2}^{(2)} \sigma_{y}-i d_{3}^{(2)} \sigma_{z},
\end{align}
where $\sigma_{x,y,z}$ are Pauli matrices, $I$ is a two-by-two identity matrix, and the coefficients $d_{i}^{(1,2)}$ ($i=0,1,2,3$) are
\begin{align}
d_{0}^{(1)}&=-\cos \tilde{k} \sin \theta_{1} \sin \theta_{2}+\cos \theta_{1} \cos \theta_{2}, \nonumber\\
d_{1}^{(1)}&=-\cos \theta_{2} \sin \theta_{1}-\cos \tilde{k} \cos \theta_{1} \sin \theta_{2}, \nonumber\\
d_{2}^{(1)}&=-\sin \tilde{k} \sin \theta_{2}, \nonumber\\
d_{3}^{(1)}&=0,
\end{align}
and
\begin{align}
d_{0}^{(2)}&=-\cos \tilde{k} \sin \theta_{1} \sin \theta_{2}+\cos \theta_{1} \cos \theta_{2}, \nonumber\\
d_{1}^{(2)}&=\cos ^{2} \tilde{k} \sin \theta_{1}\left(1-\cos \theta_{2}\right)-\sin \theta_{1}-\cos \tilde{k} \cos \theta_{1} \sin \theta_{2}, \nonumber\\
d_{2}^{(2)}&=-\sin \tilde{k}\left(\cos \theta_{1} \sin \theta_{2}-\cos \tilde{k} \sin \theta_{1}+\cos \tilde{k} \cos \theta_{2} \sin \theta_{1}\right),\nonumber \\
d_{3}^{(2)}&=0.
\end{align}
Here $\tilde{k}$ belongs to the first Brillouin zone of the momentum lattice.

The winding numbers are calculated as
\begin{equation}
C_{\alpha}=\frac{1}{2 \pi} \int d \tilde{k} \frac{-d_{2}^{(\alpha)} \frac{\partial d_{1}^{(\alpha)}}{\partial \tilde{k}}+d_{1} \frac{\partial d_{2}^{(\alpha)}}{\partial \tilde{k}}}{{d_{2}^{{(\alpha)}_2}}+{d_{1}^{{(\alpha)}_2}}} \quad (\alpha=1,2).
\end{equation}

Alternatively, we define the topological invariants
\begin{align}
C_{0}&=\frac{C_{1}+C_{2}}{2} ,\nonumber\\
C_{\pi}&=\frac{C_{1}-C_{2}}{2},
\end{align}
which directly correspond to topological edge states with quasienergies $\epsilon=0$ and $\epsilon=\pi$, respectively, through the bulk-boundary correspondence. The topological phase diagram for $(C_0,C_\pi)$ are shown in Fig.~3 of the main text.

\subsection{Time averaged mean chiral displacement}
Topological invariants can be directly probed in quantum-walk dynamics through mean chiral displacement (MCD), defined as
\begin{align}
\mathcal{C}_{1,2}(t)=2\sum_{m}m[P^{(1,2)}_{m,a}(N_{\rm step})-P^{(1,2)}_{m,b}(N_{\rm step})],
\end{align}
where $P^{\alpha}_{m,a(b)}(N_{\rm step})$ is the atom population in $|m,a(b)\rangle$ at the $N_{\rm step}$-th step under $U_\alpha$. In our experiment, we measure the time-averaged mean chiral displacement (A-MCD) instead, which converges faster to the corresponding winding number. The A-MCD is defined as
\begin{align}
\bar{\mathcal{C}}_{1,2}(t)=\frac{2}{N}\sum_{N_{\rm step},m}m[P^{(1,2)}_{m,a}(N_{\rm step})-P^{(1,2)}_{m,b}(N_{\rm step})],
\end{align}
where $N_{\rm step}=1,2,\cdots, N$.

\begin{figure}[tbp]
\includegraphics[width=0.35\textwidth]{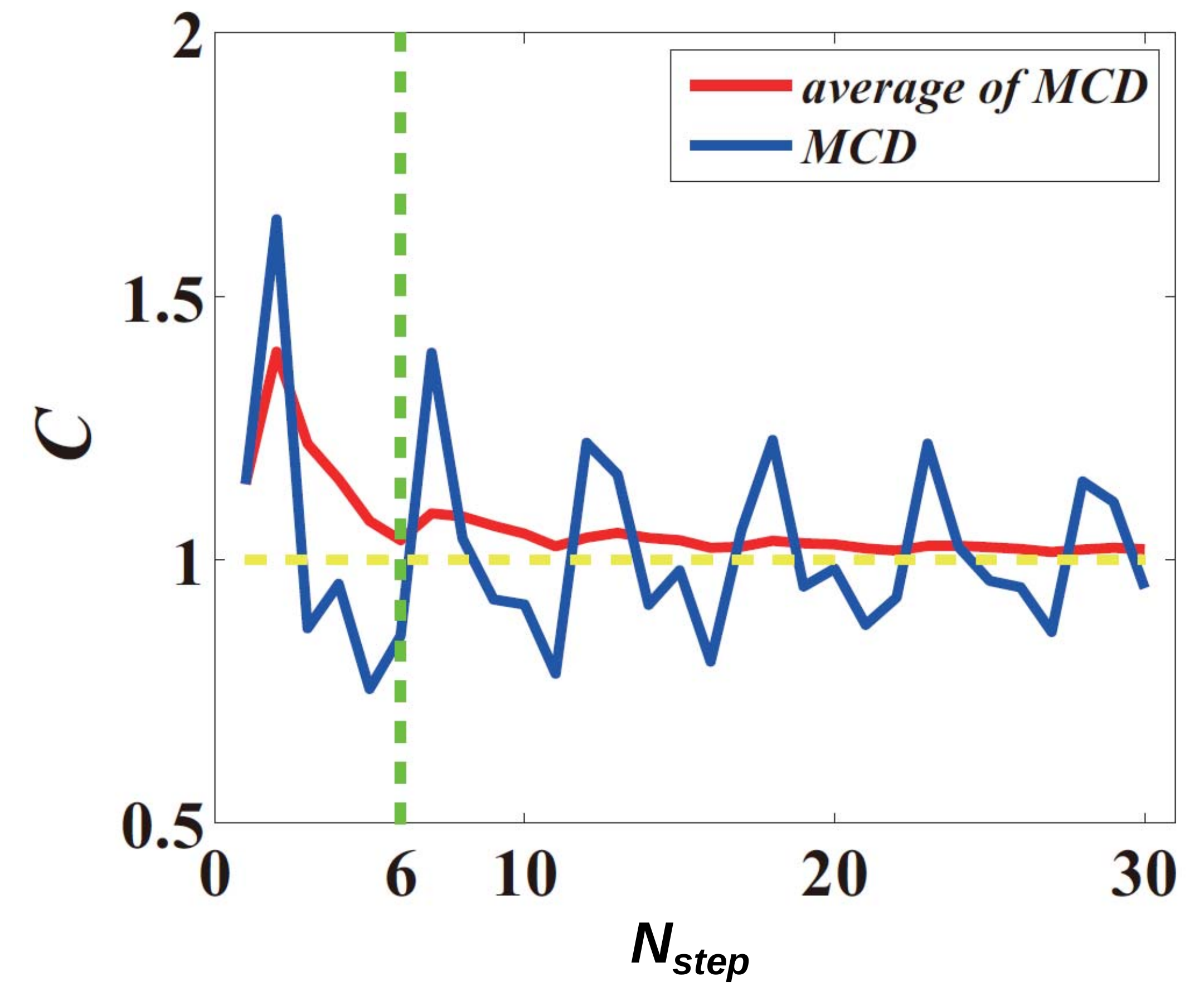}
\caption{Comparison of MCD and A-MCD under the effective Hamiltonian simulation. The blue line shows MCD versus discrete time step $t/T$. The red line shows A-MCD, which converges faster than the MCD. The yellow dashed line shows the ideal winding number $C_1$. The parameters are $(\theta_1, \theta_2)=(3\pi/32,9\pi/32)$. The green dashed line indicates the experimentally implemented time step $N_{\rm step}=6$.} \label{SMfig:MCD}
\end{figure}

In Fig.~\ref{SMfig:MCD}, we show a numerical calculation comparing MCD and A-MCD for a finite-step quantum walk governed by the effective Hamiltonian Eq.~(\ref{eq:s8}).
Whereas oscillations in the MCD persist into longer times, A-MCD already converges at the sixth step. Apparently, the fast convergence of A-MCD stems from the oscillatory behavior of MCD at intermediate times. Further, from an experimental point of view, A-MCD helps to average out
the inevitable fluctuations in the durations of Raman pulses.
As such, it is preferable to probe A-MCD for the direct measurement of winding numbers.

\subsection{Detecting of edge states}

\begin{figure*}[tbp]
\includegraphics[width= 0.6\textwidth]{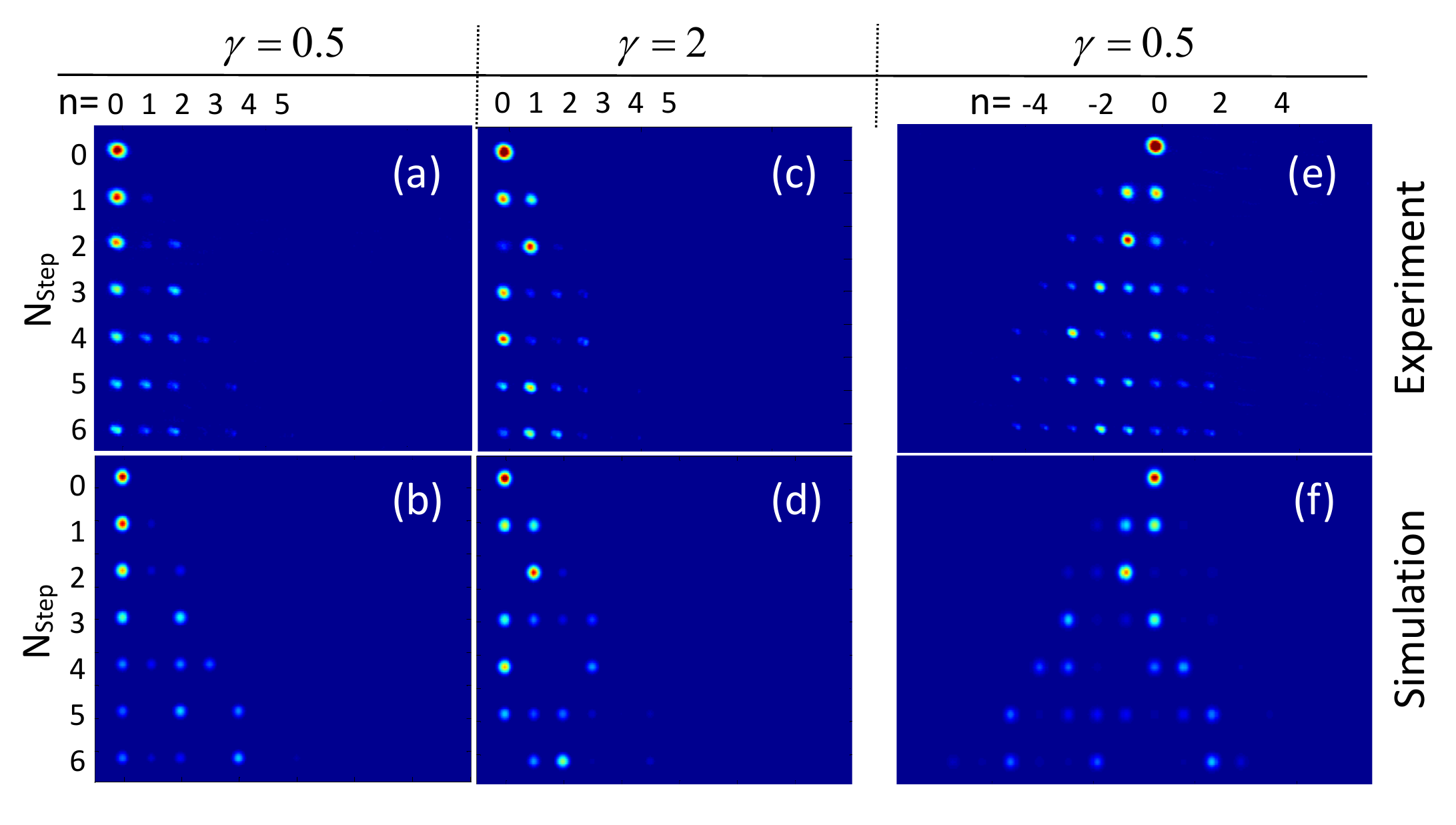}
\caption{
Experimental detection of topological edge states. (a)(b) Dynamics under $U_1$ with an open boundary at $n=0$. The bulk is topologically non-trivial with $\gamma=0.5$ ($\gamma=\theta_1/\theta_2$) and $C_1=1$. (c)(d) Dynamics under $U_1$ with an open boundary at $n=0$. The bulk is topologically trivial with $\gamma=2$ and $C_1=0$. Localization of atom population near $n=0$ is observed in (a)(b), but not in (c)(d), particularly for the last step. (e)(f) Dynamics under $U_1$ with no boundaries near $n=0$. The bulk is topologically non-trivial with $\gamma=0.5$ and $C_1=1$. Compared to (a)(b), atoms are not localized near $n=0$. The simulaitons are done with the full Hamiltonian.
}\label{fig:SMedge}
\end{figure*}

\begin{figure}[b]
\includegraphics[width=0.8\textwidth]{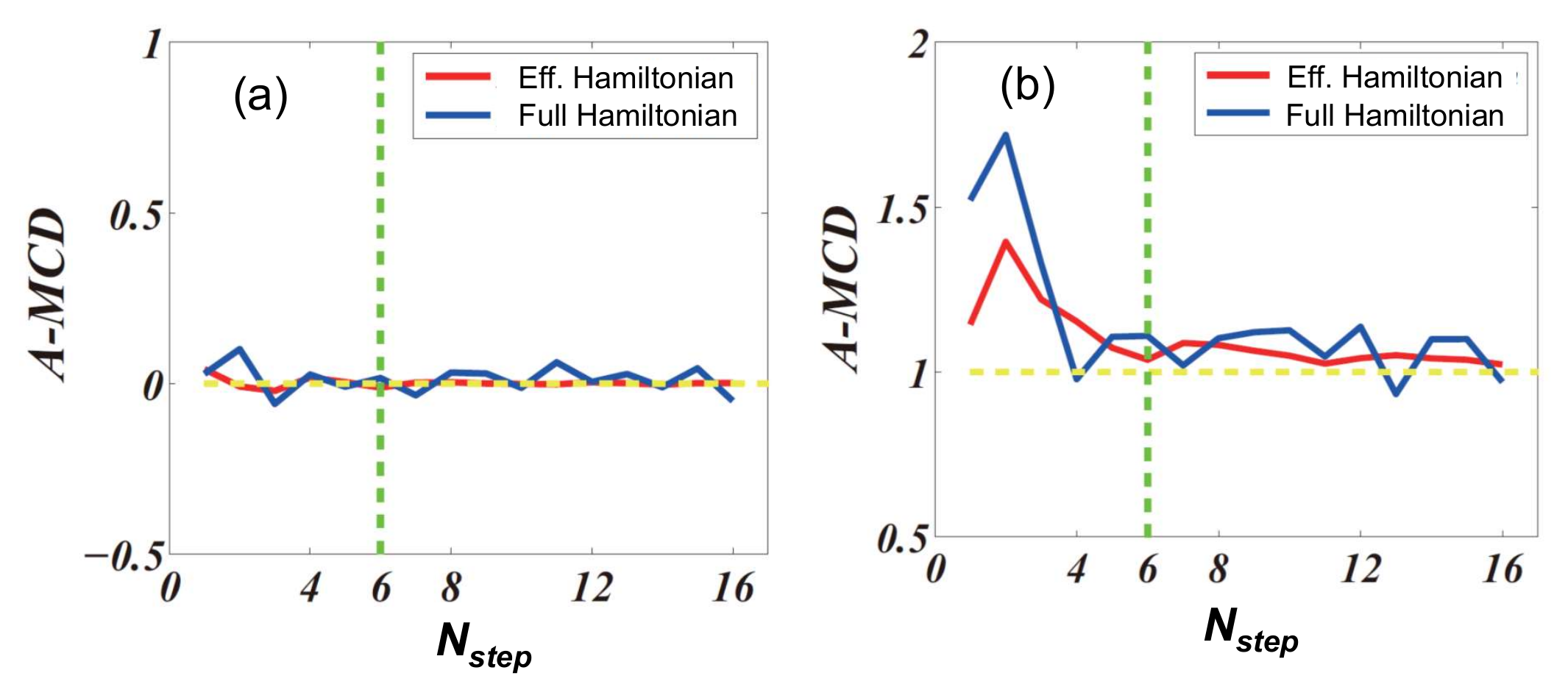}
\caption{A-MCD for (a) $(\theta_1,\theta_2)=(5\pi/16,\pi/16)$, and (b) $(\theta_1, \theta_2)=(3\pi/32,9\pi/32)$. The red line shows A-MCD under the effective Hamiltonian Eq.~(\ref{eq:s8}). The blue line shows A-MCD of the full Hamiltonian Eq.~(\ref{eq:Heff}). The green dashed line indicates the experimentally implemented time step $N_{\rm step}=6$.}\label{SMfig:com}
\end{figure}

A prominent property of topological phases, robust edge states exist at boundaries between bulks with different topological invariants. The presence of edge states give rise to local population accumulation in quantum-walk dynamics, which have been used as a signature for topological quantum walks. For our system, we create an open boundary near $n=0$ by turning off RF components in the AOM dirver corresponding to sites with $n<0$. In Fig.~\ref{fig:SMedge}, we show measured atom population along the momentum lattice at different times of the dynamics. Under an open boundary condition, local atom population should accumulate near $n=0$ when the bulk has finite winding numbers. In contrast, when the bulk winding numbers vanish, atom population should become extended in momentum space at long times. As illustrated in Fig.~\ref{fig:SMedge}(a)(c), difference in the local-population accumulation near the boundary for topological trivial and non-trivial bulks is becoming discernable at the sixth step,  albeit larger steps are needed to fully differentiate the two cases.
For comparison, we plot in Fig.~\ref{fig:SMedge}(e)(f) dynamics of a homogeneous quantum walk. Despite finite winding numbers in the bulk, atom population becomes extended at long times, due to the absence of any boundaries.

\subsection{\label{sec:level1}Comparsion of the full Hamiltonian and the effective Hamiltonian}
In this section, we compare dynamics under the full Hamiltonian in Eq.~(\ref{eq:Heff}) and that under the simplified periodically driven SSH model Eq.~(\ref{eq:s8}). Using numerical simulations, we demonstrate that, under our typical experimental conditions, it is reasonable to neglect the non-resonant terms in Eq.~(\ref{eq:Heff}), a crucial approximation leading to the implementation of discrete-time quantum-walk dynamics in momentum space.

We adopt the time frame governed by $U_1$, and choose two sets of parameters: $(\theta_1,\theta_2)=(5\pi/16,\pi/16)$ with $C_1=0$, and $(\theta_1,  \theta_2)=(3\pi/32,9\pi/32)$ with $C_1=1$.
In Fig.~\ref{SMfig:com}, we plot the resulting A-MCD under the ideal Hamiltonian in red, and the A-MCD under the full Hamiltonian in blue.
In both cases, the two results lie close to one another and to the respective winding number $C_1$ (yellow dashed line) up to sixteen time steps.

\subsection{Derivation of Eq.~(5) and the impact of interactions}

In this section, we analyze the impact of interactions on quantum-walk dynamics, and derive Eq.~(5) in the main text. Since the momentum-lattice sites are typically macroscopically populated in our experiment, we take the Hartree-Fock approximation and write the interacting Hamiltonian as
 \begin{align}
H_{\rm int}=\frac{g}{V}\sum_{n\neq m}a_{n}^{\dagger}a_{m}^{\dagger}a_{n}a_{m}+\frac{g}{V}\sum_{n\neq m}b_{n}^{\dagger}b_{m}^{\dagger}b_{m}b_{n}+\frac{2g}{V}\sum_{n,m}a_{n}^{\dagger}b_{m}^{\dagger}b_{m}a_{n}+\frac{g}{2V}\sum_{n}a_{n}^{\dagger}a_{n}^{\dagger}a_{n}a_{n}+\frac{g}{2V}\sum_{n}b_{n}^{\dagger}b_{n}^{\dagger}b_{n}b_{n},
\end{align}
where $a^\dag_m$ ($b^\dag_m$) creates an atom in the state $|m,a\rangle$ ($|m,b\rangle$), $V$ is the volume of the system, and $g=4\pi\hbar^2 a_s/\mu$ with $a_s$ the $s$-wave scattering length and $\mu$ is the atomic mass.

The system's Hamiltonian is then $H=H_0+H_{\rm int}$, with
\begin{align}
\label{H02}
H_{0}=w(t)\sum_{m}b_{m}^{\dagger}a_{m}+q(t)\sum_{m}a_{m+1}^{\dagger}b_{m}+H.c..
\end{align}
Applying the Heisenberg equation, and taking the mean-field approximation $\phi_{m,a}=\frac{1}{\sqrt{N_t}}\langle a_{m}\rangle$($\phi_{m,b}=\frac{1}{\sqrt{N_t}}\langle b_{m}\rangle$), we have the equations of motion
\begin{align}
\label{CPeq}
i\hbar\frac{d}{dt}\Phi=\bar{H}\Phi,
\end{align}
with $\Phi=[\phi_{m,a},\phi_{m,b}]^{T}$ and the matrix elements of $\bar{H}$ are ($\sigma=a,b$)
\begin{align}
&\bar{H}_{m,a;m,b}=\bar{H}_{m,b;m,a}=w(t)\\
&\bar{H}_{m+1,a;m,b}=\bar{H}_{m,b;m+1,a}=q(t)\\
&\bar{H}_{m,\sigma;m,\sigma}=U|\phi_{m,\sigma}|^{2}+2U\sum_{m'\neq m,\sigma'}|\phi_{m',\sigma'}|^{2}.
\end{align}
Here $N_t$ is the total particle number, $U=g\rho$, and $\rho$ is the BEC density.

\begin{figure*}[tbp]
\includegraphics[width= 0.98\textwidth]{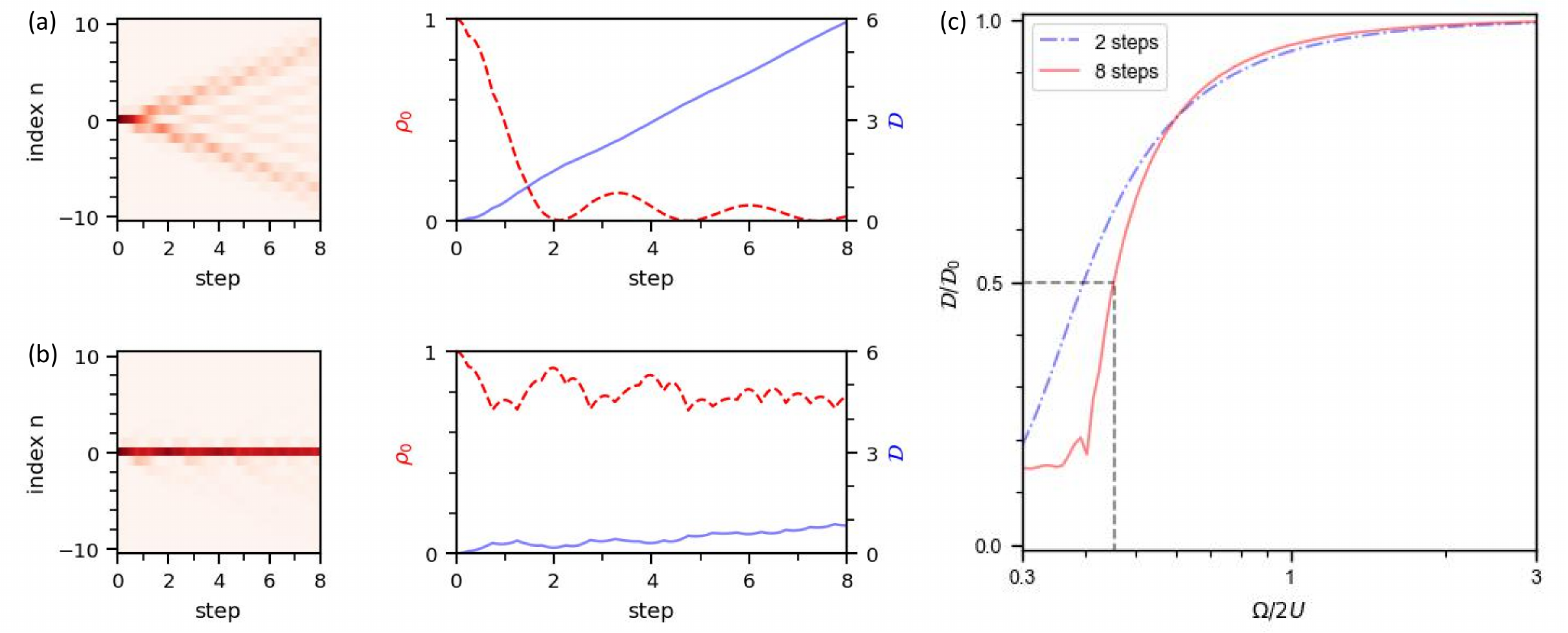}
\caption{Numerical results on the effect of interactions. For numerical calculations, we choose $\theta_1=\theta_2=3\pi/16$. (a)(b) Left: numerically simulated atom population along the lattice at different time steps, with (a) $\hbar\Omega/2U=2$ and (b) $\hbar\Omega/2U = 0.25$. Right: density at the initial site $\rho_0$ (red) and the mean distance $\mathcal{D}$ during the quantum-walk dynamics.
(c) The normailized mean distance $\mathcal{D}/\mathcal{D}_0$ as a function of $\hbar\Omega/2U$ for finite-step quantum walks.
}\label{figs4}
\end{figure*}

To show the impact of interactions, we evolve the system according to the coupled equations (\ref{CPeq}) with the atoms initialized in the $|m=0,a\rangle$ (i.e., $n=0$) state. As shown in Fig.~\ref{figs4}(a), when $U\sim\hbar\Omega$, the interaction potential is small, and its impact on the quantum-walk dynamics is not apparent.
The mean distance from the initial position
\begin{align}
\mathcal{D}=\sum_m (|2m||\phi_{m,a}|^2+|2m+1||\phi_{m,b}|^2),
\end{align}
appears to be linear in the number of steps; and there is a rapid decay in the population of the initial site. In contrast, for $U/\hbar=2\Omega$, the interaction potential is large enough to significantly affect the quantum-walk dynamics. As shown in Fig.~\ref{figs4}(b), the atoms now become localized near the initial lattice site.

In Fig.~\ref{figs4}(c), we numerically calculate $\mathcal{D}$ with increasing $\hbar\Omega/2U$, and for finite-step quantum walks with total number of steps $N_{\rm step}=8$ and $N_{\rm step}=2$ respectively. In the case of $N_{\rm step}=8$, a localization-delocalization transition can be identified near $\hbar\Omega/2U\sim 0.45$, and $\mathcal{D}/\mathcal{D}_0$ saturates to 1 for $\hbar\Omega/2U>1$. In the case of $N_{\rm step}=2$, the behavior is similar, but the transition is more smooth.

\clearpage
\end{widetext}

\end{document}